\documentstyle[aps,epsf,twocolumn,floats,prb]{revtex}
\begin{document}
\parindent=0pt
\def \be{\begin{equation}}
\def \ee{\end{equation}}
\def \bea{\begin{eqnarray}}
\def \eea{\end{eqnarray}}
\def \bpi{{\vec{\pi}}}
\def \bn{{\bf n}}
\def \br{{\bf r}}
\def \bx{{\bf x}}
\def \bk{{\bf k}}
\def \bq{{\bf q}}
\def \bQ{{\bf Q}}
\def \bS{{\bf S}}
\def \bA{{\bf A}}
\def \bL{{\bf L}}
\def \bhn{{\bf \hat n}}
\def \cH{{\cal H }}
\def \cA{{\cal A }}
\def \cL{{\cal L }}
\def \cB{{\cal B }}
\def \cS{{\cal S }}
\def \cP{{\cal P}}
\def \cO{{\cal O}}
\def \cM{{\cal M}}
\def\a{{\alpha}}
\def\b{{\beta}}
\def\d{{\delta}}
\def\e{{\epsilon}}
\def\s{{\sigma}}
\def\t{{\theta}}
\def\w{{\omega}}
\def\W{{\Omega}}
\def\D{{\Delta}}
\def\tp{{t^\prime}}
\def\eg{{\it e.g.\/}}
\def\ie{{\it i.e.\/}}
\def\etal{{\it et al.\/}}
\def\ket#1{{\,|\,#1\,\rangle\,}}
\def\bra#1{{\,\langle\,#1\,|\,}}
\def\braket#1#2{{\,\langle\,#1\,|\,#2\,\rangle\,}}
\def\expect#1#2#3{{\,\langle\,#1\,|\,#2\,|\,#3\,\rangle\,}}
\def \half{{1\over 2}}
\def\nd{{^{\vphantom{\dagger}}}}
\def\yd{^\dagger}
\def \yBCO6{{ YBa$_2$\-Cu$_3$\-O$_{7-\delta}$ }}
\def \yBCO6{{ YBa$_2$\-Cu$_3$\-O$_{6.6}$ }}
\def \yBCO6x{{YBa$_2$\-Cu$_3$\-O$_{6+x}$}}
\def \yBCFO{{YBa$_2$Cu\-$_{2.55}$\-Fe$_{0.45}$\-O$_{y}$}}
\draft
\twocolumn[\hsize\textwidth\columnwidth\hsize\csname @twocolumnfalse\endcsname

\title{Plaquette Boson-Fermion Model of Cuprates}
\author{Ehud Altman and Assa Auerbach}
\address{ Department of
Physics, Technion, Haifa 32000, Israel.} \date{\today} \maketitle
\begin{abstract}
The strongly interacting Hubbard model on the square lattice is
reduced to the low energy Plaquette Boson Fermion Model (PBFM).
The four bosons  (an antiferromagnon triplet and a d-wave hole pair), and the fermions are defined
by the lowest plaquette eigenstates. We apply the Contractor Renormalization method of Morningstar and
Weinstein to compute the boson effective interactions. The
range-3 truncation error is found to be very small, signaling short hole-pair  and magnon
coherence lengths. The pair-hopping and magnon interactions are comparable, which explains
the rapid destruction of antiferromagnetic order with emergence of superconductivity, and validates
a key assumption of the projected SO(5) theory.
A vacuum crossing at larger doping marks a transition into the overdoped regime.
With hole fermions occupying small Fermi pockets and Andreev coupled to hole pair bosons,
the PBFM yields several testable predictions for photoemmission, tunneling asymmetry
and entropy measurements.
\end{abstract}
\vskip2pc]
\narrowtext
\section{Introduction}\label{intro}
 In 1987,  shortly after the discovery of  high temperature
superconductivity in cuprates, Anderson\cite{rvb} proposed that
the key to this perplexing  phenomenon hides in the large positive
Hubbard   interactions in the copper oxide planes. Indeed, at zero
hole doping,   the Hubbard model  captures the  Mott insulator
physics of the parent compounds  e.g. La$_2$CuO$_4$. The doped
Hubbard model  however has so far resisted a  definitive solution,
primarily because its spins and holes  are highly entangled with
no obvious small parameter to separate them. Whether the  Hubbard
model even supports superconductivity without additional
interactions remains a subject of controversy. Different mean
field theories suggest conflicting ground state order parameters
and correlations. Numerical methods are restricted to finite
clusters where hole pairing is found\cite{PB,WS,dag}, but
off-diagonal long range order has not been ascertained.

\begin{figure}[htb]
\begin{center}
\vspace*{13pt}
 \leavevmode \epsfxsize0.75\columnwidth
\epsfbox{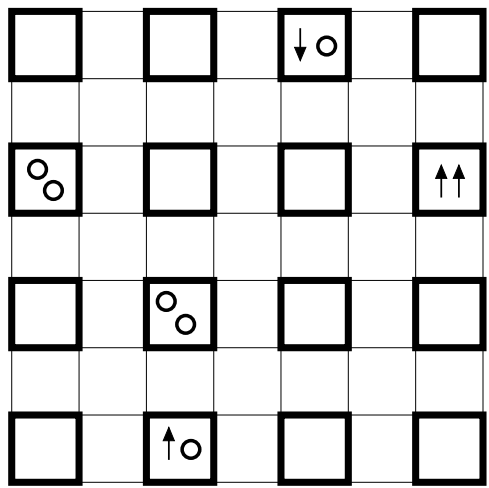}\vskip1.0pc \caption{{\bf Local bosons and
fermions  on the plaquette lattice.} The singlet RVB  vacua are
depicted as solid squares. Holes are depicted by circles. The triplets, single hole  and hole
pairs Hubbard eigenstates define the degrees of freedom of the
effective Plaquette Boson-Fermion Model. Interplaquette couplings
are computed using Contractor Renormalization. }
\label{fig:scheme}
\end{center}
\end{figure}

This paper charts a route from  the microscopic Hubbard model on
the square lattice to  an effective lower energy Plaquette Boson
Fermion Model (PBFM) at low hole doping. We apply the Contractor
Renormalization (CORE) method of Morningstar and
Weinstein\cite{core} to the plaquettized lattice (see
Fig.\ref{fig:scheme}) in a one step transformation.

We find that the  bosonic part of the effective Hamiltonian is
closely related to the  projected SO(5) (pSO(5))
theory\cite{pSO5,pSO5-num}; a theory of four bosons: a hole pair
and a (antiferro)magnon triplet. The pSO(5) model describes the
competition between antiferromagnetism
 and superconductivity in a quantum mechanical framework.
Its mean field theory yields some broad features of the cuprate
 phase diagram at low doping. At low temperature, the hole pairs  are governed by
a phase fluctations action\cite{Uemura,EmeryKivelson} which
explains the (non BCS) proportionality between superfluid density, transition temperature
and hole concentration. In the superconducting phase, the
magnons are massive and give rise to an antiferromagnetic resonance
in neutron scattering. These massive magnons were argued to produce
a resistance peak series in Josephson junctions\cite{prl}.

Nevertheless  without a microscopic foundation, the fundamental
``mechanism'' problem remains: What creates and holds together
d-wave hole pairs in the presence of local repulsive interactions,
without the benefit of retardation and phonons? Even assuming that
hole pairs move coherently, what is their hopping rate, and is it
of the same order as the Heisenberg exchange energy as assumed by
the pSO(5) theory?

Here we address these questions and afford the pSO(5)theory and
its phase diagram a microscopic foundation. We also include the
hole fermions which provide gapless (nodal)
excitations in the square lattice. Their bandstructure is obtained
from previously published numerical results, and their coupling to
the bosons is estimated by symmetry and microscopic
considerations.

The resulting Plaquette Boson Fermion Model (PBFM) describes two
coupled charged systems (i) Hole pair bosons, which Bose condense below $T_c$  and induce a proximity
gap on the holes. (ii) The hole fermions, which occupy small Fermi pockets around
$(\pm\pi/2,\pm\pi/2)$ and have a large
van-Hove peak in density of states near $(\pm\pi,0),(0,\pm\pi)$,
the ``antinodal'' points.

We discuss the thermodynamics of the coupled system, with the
constraint on the total doping density. Previously proposed
boson-fermion models\cite{bosonfermion} differ from the PBFM by
their Hilbert space  (e.g. by having a large Fermi surface, and
counting occupations from the electron vacuum).

For the PBFM, in the weak coupling approximation,  some straightforward experimental
implications  are obtained:

\begin{enumerate}
\item {\em Hole spectral weight in Luttinger theorem-violating
momenta} (outside the ``large'' electron Fermi surface), e.g. on
the line $(\pi,0) \to (\pi,\pi)$\cite{pi0-pipi}. This weight
survives above $T_c$ and is  associated with excited holes moving
in the correlated RVB vacuum.
\item
{\em Asymmetry  in tunneling conductance.}  At low doping, $x<<1$,
particle-hole symmetry is expected to be violated i.e.
the positive bias conductance (injection of
electrons) is suppressed by a factor proportional to $x$, relative
to the negative bias conductance (injection of holes). Such a trend indeed appears in
tunneling data\cite{tunneling}.

\item {\em The pseudogap doping dependence.}
The pseudogap energy in photoemmission\cite{ZX,campuzano} and
tunneling\cite{tunneling}
is at the  van-Hove peak of antinodal fermions. The decrease of
pseudogap with doping follows the increase in fermion chemical
potential. Its derivative with respect to doping measures the
combined hole fermions' and hole pair bosons' compressibilities.
\item {\em Nodal transverse velocity.}  The quasiparticles proximity  gap
near the nodal directions determines their transverse velocity.
This velocity can be measured by photoemmission and optical
conductivity\cite{CP}. We expect it to vanish  at $T_c$, and to be
proportional to the Bose condensate order parameter. Thus it
should scale as $ v_\perp \propto \sqrt{T_c(x)}$.
\item {\em  Hole dependent entropy.} At temperatures above the superconducting
transition,
hole pairs evaporate
into hole fermions, because of the difference in
their density of states. The doping dependent
entropy\cite{loram} is dominated by the fermion  contribution.
\end{enumerate}

The paper is organized as follows: In Section II we introduce the
eigenstates of the  Hubbard model on a plaquette. The local bosons
and fermions  are defined as the creation operators of these
eigenstates. The physics learned from the four site problem is
instructive: Undoped, the ground state is a local resonating
valence bonds  (or projected $d$-wave BCS) state. The four bosons
create the lowest triplet and the hole pair singlet. There are two
degenerate spin half  plaquette fermion states with symmetry
$(\pi,0)$ and $(0,\pi)$.

It has long been appreciated  that in the Hubbard model,  two
holes cannot bind on a dimer bond, but they can bind on a
plaquette (and on larger clusters)\cite{pairbinding}. The hole
pair wave function has $d_{x^2-y^2}$ symmetry for $\pi/2$
rotations. The next step is to compute their interplaquette
hopping rate in order to see whether they can preserve their
integrity  on the infinite square lattice. A short discussion is
included about plaquette ``vacuum crossing'', which occurs at
large chemical potentials and may be associated with a transition
from underdoped to the overdoped regime.

 In Section III
the Contractor Renormalization (CORE)   method is reviewed.  The
method requires exact diagonalization of  multi-plaquette
clusters, in principle up to infinite range. Of course, the method
is useful only if it converges rapidly in a feasible range of
interactions. We have tested the convergence of the  low spectrum
for the Hubbard models on open ladders, with satisfying results.
These tests confirm our belief that the convergence depends on a
short boson coherence length, of order one plaquette size. This is
very encouraging for the useful application of CORE to our
problem, since the experimental superconducting coherence length
of cuprates  also appears to be particularly short in the
underdoped systems. We discuss the artifacts of the {\em formal}
translational  symmetry breaking within  CORE. In
Appendix~\ref{app:tightbinding}, we use the tight binding model as
a pedagogical example of how longer range interactions of CORE
serve to restore an unphysically broken symmetry.

In Section IV  the Plaquette Boson-Fermion model is derived. We
discuss the hole pairs integrity, as evidenced from the numerical
results, and how it is related to  the sizeable pair hopping
energy. The pair kinetic energy is crucial in stabilizing
superconductivity. The full four boson hamiltonian is given in
Appendix \ref{app:4BM}.  The hole fermions band structure and interactions
with the bosons are added. The thermodynamics of the weakly
coupled PBFM yields a relation between the pseudogap energy, the
bosons and fermions compressibilities, and the evaporation of hole
pairs into fermions at higher temperature.

We conclude with a summary and a discussion of future directions in Section V.

\section{Plaquette States}
We study the Hubbard model
\be
\cH=-t\sum_{\langle i j\rangle,s}^{sl}
\left( c^\dagger_{is} c^\nd_{js} + \mbox{H.c}  \right)   +U\sum_i
 n_{i\uparrow} n_{i\downarrow} , \label{HM}
 \ee
 where $c^\dagger_{is}, n_{i s}$ are electron creation and number operators at site $i$ on the
 square lattice.
 We will occasionally refer to its Gutzwiller projected version, the  t-J model:
 \be
 \cH^{tJ}=   -t\cP  \sum_{\langle i j\rangle,s}   \left( c^\dagger_{is}
c^\nd_{js} + \mbox{H.c}  \right)\cP    + J \sum_{\langle ij\rangle}
 \bS_{i} \cdot \bS_{j} + \cal{J}'.
\label{tJ}
 \ee
 $\cP$ is the projector of doubly occupied
 states, and $J\to 4t^2/U$ at large $U/t$. $\cal{J}'$ is the term of order $J$
 that includes next nearest neighbor hole hopping \cite{AssaBook3}.
 For one electron per site (half filling), the short range antiferromagnetic
 correlations are apparent when diagonalizing (\ref{HM}) and (\ref{tJ})  on two sites.
The dimer states were used to construct effective models on the
ladder\cite{gopalan,gayen} and for spin-Peierls phases on the
square lattice\cite{sachdev}. The projected SO(5) model was
defined on a ladder using empty dimer states as the hole pair
bosons\cite{pSO5}. However, there is {\em no hole pair binding}
for the Hubbard model on a  dimer. Naively, this suggests that
pairs could readily disintegrate into single holes once
inter-dimer hopping is turned on.
Moreover, if one wishes to capture $d$-wave symmetry in the hole
pair wavefunction, the basic unit block must possess at least
four-fold rotational symmetry.

The smallest such block that can cover the square lattice is the four site plaquette.
It is  a trivial task to diagonalize the Hubbard model on a plaquette and
obtain its spectrum and wavefunctions.

\begin{figure}[htb]
\begin{center}
\vspace*{13pt}
 \leavevmode \epsfxsize1.0\columnwidth
\epsfbox{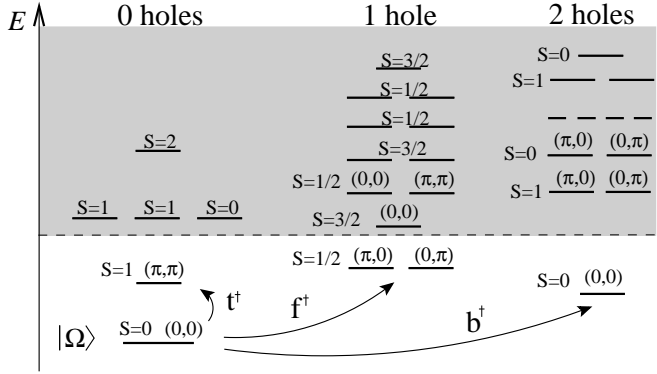}\vskip0.8pc
\caption{{\bf Lowest spectrum of the Hubbard model on a plaquette}.  Eigenstates are labelled
by total spin $S$ and plaquette momentum $Q_x,Q_y=0,\pi$. The shaded area
is over all high energy truncated states. The vacuum is defined as $\ket\W$, and the second
quantized operators connect the vacuum to the lowest eigenstates as shown.
}\label{fig:spectrum}
\end{center}
\end{figure}

The spectrum is depicted in Fig.~\ref{fig:spectrum}.

Since  it is cumbersome to write the full wavefunctions explicitly, we represent  their dominant correlations as follows
(i)  Real space (RS) description using  holes,  dimer singlets and dimer
triplets  as depicted in  Fig.~\ref{fig:RS}.
(ii) Plaquette momenta  (PM) representations using
$\bQ=(Q_x,Q_y)$, $Q_\alpha =0,\pi$,   the four points on the plaquette Brillouin zone. The plaquette electron operator
of plaquette $i$ is given by
\be
c^\dagger_{\bQ is}= {1\over 2}
\sum_{\eta= 0, \hat{x}, \hat{y},\hat{x}+\hat{y}} e^{i\bQ \cdot \eta} c^\dagger_{i+\eta s}.
\ee

It is instructive to examine
the plaquette eigenstates and energies in some detail before proceeding to couple them.

 \subsection{The vacuum}
The ground state of the 4-site Hubbard model  at half filling ($n_e=4$)  is called $\ket\W$.
In the PM representation  it can be described by (suppressing the plaquette index),
\be
\ket\W = {\cP\over \sqrt{Z_\Omega}}  (c^\dagger_{(\pi,0)\uparrow} c^\dagger_{(\pi,0)\downarrow}   -
c^\dagger_{(0,\pi)\uparrow}  c^\dagger_{(0,\pi)\downarrow} )
c^\dagger_{(0,0)\uparrow} c^\dagger_{(0,0)\downarrow} |0\rangle.
\label{W}
\ee
$Z$ is the wavefunction normalization factor.
 $\ket\W$ is  a  d-wave BCS state, where doubly occupied states are suppressed by
 a partial  Gutzwiller projection   $\cP(U/t)$. (At large $U$, $\cP$  becomes a full projection).

\begin{figure}[htb]
\begin{center}
\vspace*{10pt}
 \leavevmode \epsfxsize0.75\columnwidth
\epsfbox{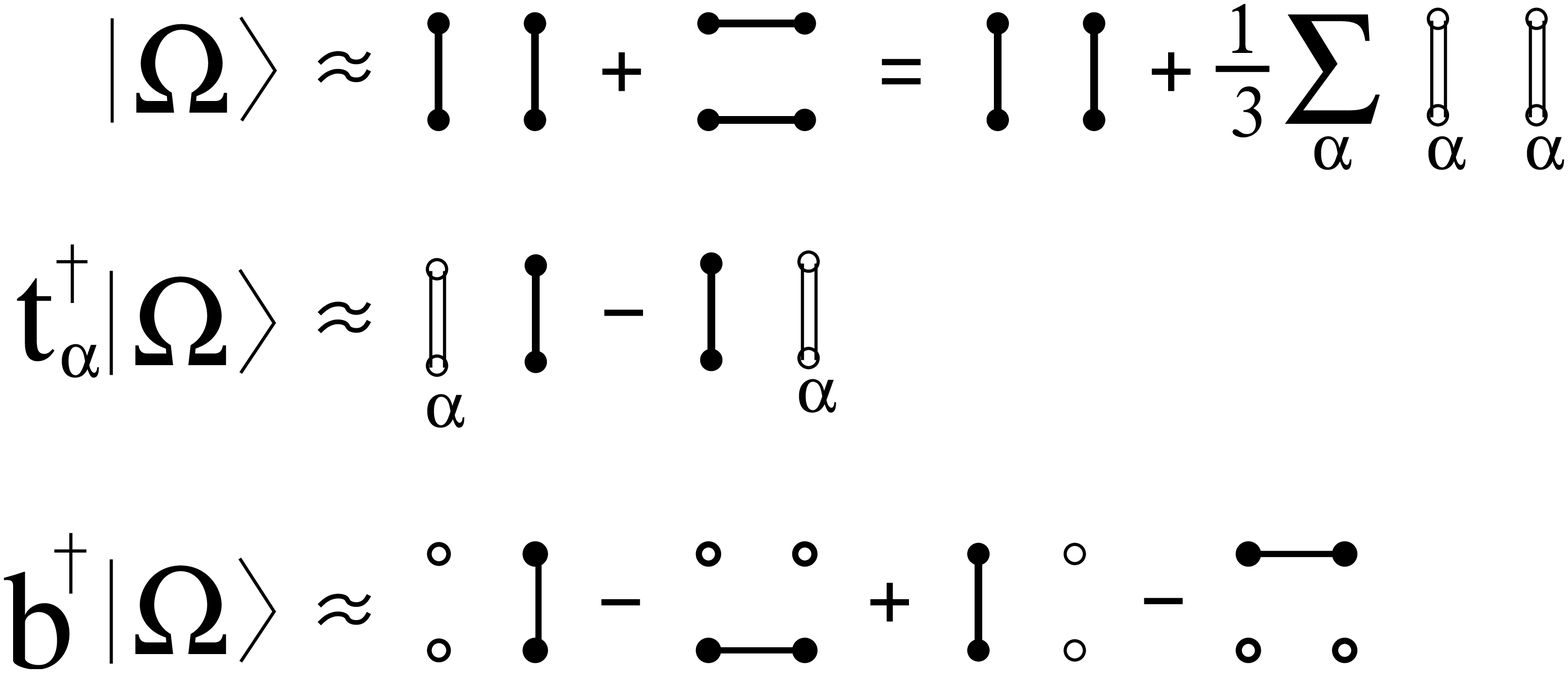}\vskip1.0pc
\caption{{\bf Real space representation of plaquette bosons}.  Dominant spin
and charge correlations in the plaquette
bosons wavefunctions.  Bold lines represent singlet dimers
$ \uparrow_i \downarrow_j  -  \downarrow_i  \downarrow_j $,
and double lines represent the triplet
$ \uparrow_i \downarrow_j  +  \downarrow_i  \uparrow_j $,
$ \uparrow_i \uparrow_j  \pm   \downarrow_i  \downarrow_j $, where $i$ and $j$ are on sublattices $A$ and $B$
respectively.
Holes are depicted by open circles.
}\label{fig:RS}
\end{center}
\end{figure}

In the RS representation, see Fig.\ref{fig:RS},  $\ket\W$ is depicted
as the resonating valence bonds (RVB) ground state of the
Heisenberg model plus small contributions from doubly occupied
sites.   In the two dimer basis, $\ket\W$ contains a large
contribution from a triplet pair.

The product state  $\ket\W
=\prod_i^{plaq}\ket\W_i$, is our vacuum state for the full lattice, upon which Fock states
can be constructed using second quantized boson and fermion
creation operators.

\subsection{Magnon triplet}
The  magnons are defined by the lowest  triplet of $S=1$ states.
In PM representation they are \be t^\dagger_\alpha \ket\W =
{\cP\over  \sqrt{Z_t} } \sum_{\bQ s} c^\dagger_{\bQ s}
\sigma^\alpha_{ss'} c^\nd_{\bQ+(\pi,\pi) s'}
\ket\W,~~~\alpha=x,y,z ,\label{magnons} \ee
where $\sigma^\alpha$ are Pauli matrices.
These
(antiferro-)magnons have plaquette momentum $\bQ=(\pi,\pi)$.
Their excitation energy  is close to the
superexchange  energy $J \approx 4t^2/U$. An antiferromagnetic state can be constructed
by a product of plaquette coherent states
\be
\Psi^{afm}~ =\prod_i^{plaq}  (\cos\theta+   \sin\theta m^\alpha ~t^\dagger_{i
\alpha} )\ket\W, \label{Neel} \ee
where $|\bf{m}|=1$. This state supports a  finite
staggered moment
\bea
{1\over N} \langle
S^\alpha_{(\pi,\pi)}\rangle_{\theta,m^\alpha}  &=&
\sqrt{3/8} m^\alpha \cos\theta\sin\theta \nonumber\\
& \le& 0.306.
\label{stagm}
\eea
Note that the maximal magnetization per site
supported by $\Psi^{afm}$ is less than the classical value of 0.5,
since  it does not contain higher spin states  up to $S=2$.

\subsection{Single hole  fermions}
\label{SHF}
 The  ground states for a single  hole ($n_e=3$) are
two degenerate doublets   described by plaquette momenta $\bQ =
(0,\pi), (\pi,0)$:
 \be
 f^\dagger_{\bQ s} \ket\W  =
{\cP\over \sqrt{Z_\bQ} } c_{\bQ s} +\ldots   \ket\W, ~~
s=\uparrow,\downarrow, \label{plaq-fer}
\ee
where $\ldots$ represent higher order electron operators.
The  hole fermion Bloch state can be constructed as
\be
f^\dagger_{\bk+\bQ s} \ket\W  =
\sum_i^{plaq} e^{i \bk\cdot \bx_i} f^\dagger_{\bQ is} \ket\W.
\ee
For  a  lattice of {\em disconnected}
plaquettes, $f^\dagger_{(\pi,0)}$,  creates an eigenstate with a photoemmission
spectral weight given by
  \be
|\langle \W | f_{(\pi,0)s} c_{(\pi,0)s}    \ket\W |^2
=Z_{(\pi,0)},
\label{ZQ1}
 \ee
where e.g. for the t-J model, $1/4< Z_{(\pi,0)}<1/2$ is a function of $t/J$.
This weight is further renormalized by interplaquette couplings in the effective Hamiltonian.

Incidentally, there is another  degenerate pair of  doublets at
higher energy   (of order $J$) at momenta $\bQ=(0,0), (\pi,\pi)$.
It turns out that by symmetry, the $(\pi,\pi)$ state has vanishing
hole spectral weight, that is to say for all values of
$U/t$
 \be
Z_{(\pi,\pi)} = |\langle \W | f_{(\pi,\pi)s} c_{(\pi,\pi)s}
 \ket\W |^2  =0.
 \label{ZQ2}\ee
Since these states couple by interplaquette hopping to the lower
doublet, this produces an asymmetry  of the quasiparticle weight
between momenta close to $(0,0)$ and $(\pi,\pi)$. This asymmetry
may explain the difficulty in observing   ``shadow bands'', i.e.
quasiparticles on the  Fermi pockets surfaces closer to
$(\pi,\pi)$\cite{norman-nature}.

 It is interesting to note that the {\em two-fold degeneracy}
 of the   fermion doublets
 is a property of the plaquette. The four site
 Hubbard and t-J Hamiltonians happen  to commute
 with   the {\em plaquette d-density wave operator}\cite{ddw}
 \be
\hat{D} = i   \cP \sum_{s}(c\yd_{ (\pi,0)s}c\nd_{ (0,\pi)s}-c\yd_{
(0,\pi)s}c\nd_{ (\pi,0)s} ) \cP .
\ee $\hat{D}$ connects between the
doublet pairs $(\pi,0) \leftrightarrow (0,\pi)$, and    $(0,0)
\leftrightarrow(\pi,\pi)$. Thus a possible ground state of one
hole is the current carrying state
  \be
  \Psi_s= \prod_i^{plaq}   (f\yd_{(\pi,0),is}+i f\yd_{(0,\pi),is} \ket\W,
  \ee
 which  is a staggered flux (or d-density wave) state.
 For a single hole  in  $4\times 4$ periodic lattices,
 this state does not seem to be the lowest energy(see Section \ref{subsec:fer-ham}).
 However, a large susceptibility for
 such currents is expected since the hole dispersion has a valley between the
 antinodal points  $(\pi,0)$ and $(0,\pi)$, which is weakly dispersive  and
 contains a large admixture of the two plaquette fermion states.
  It is thus conceivable that the
  staggered flux combination of the plaquette fermions would be selected
  in a vortex core  or near the sample edge.

\subsection{Hole pair boson}
\begin{figure}[htb]
\begin{center}
\vspace*{10pt}
 \leavevmode \epsfxsize1.0\columnwidth
\epsfbox{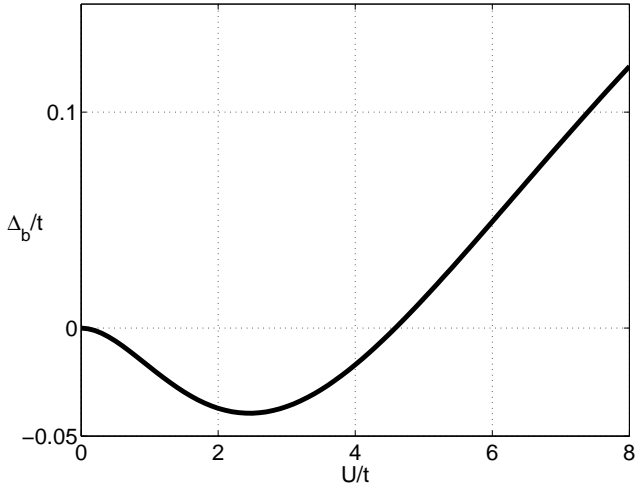}\vskip0.6pc \caption{{\bf Pair binding
energy on a plaquette}. $\Delta_{b}$ of  Eq. (\protect\ref{bind})
calculated for the Hubbard model for different interaction
strengths. For $\Delta_{b}<0 $  the  hole pair  is more stable than
two single holes on a  disconnected plaquette lattice.  }
\label{fig:pairbinding}
\end{center}
\end{figure}

The ground state of two holes ($n_e=2$)   is described by
 \bea
b^\dagger_\alpha \ket\W &=&
{1\over   \sqrt{Z_b}} \cP   c^\dagger_{(0,0) \uparrow} c^\dagger_{(0,0)\downarrow} |0\rangle
\nonumber \\
&=&   {1\over  \sqrt{Z'_b}} \left(\sum_{i j}  d_{ij}
  c_{i\uparrow}  c_{j\downarrow} +\ldots \right) \ket\W,
\label{holepairs} \eea
 where $d_{ij}$ is +1 (-1) on vertical (horizontal) bonds, and  $\ldots$ are higher order
 $U/t$-dependent operators. Thus, $b^\dagger$   creates a  pair
 with internal  $d$-wave symmetry with respect to the vacuum. For the relevant range of $U/t$, the
state normalization is $1/3<Z_b' <2/3$.
The important energy to note is the pair binding energy defined as
\be \D_{b} \equiv E(0) +E(2)-2E(1)
\label{bind} \ee
where $E(N_h)$ is the ground state of $N_h$
holes.   $\D_{b}$
is depicted in  Figure \ref{fig:pairbinding}.  In the range $U/t \in (0,5)$, it is bounded by
 $-0.04 t< \D_b<0$.   It has been well appreciated that the Hubbard,  t-J and even
 CuO$_2$ models
 have pair binding in finite  clusters starting with one plaquette\cite{pairbinding}.
 In larger clusters, such as the $4\times 4$ lattice, pair binding is seen for up to  6 holes\cite{dag} (three hole pairs)
 for  $U/t \le 20$.  This does not yet explain the integrity of pair correlations on the infinite lattice
 since the electron hopping  energy  $t$ is much larger than the pair binding energy.
 In Section \ref{sec:mechanism} and Appendix \ref{app:4BM},  we  show numerical evidence that
 plaquette pairs
 {\em survive}  disintegration into fermions.

A d-wave superconducting state  can be written as the coherent state
\be
\Psi^{d-scF}\equiv \prod_i^{plaq} (\cos\theta +   \sin\theta e^{i\varphi}  b^\dagger_i)\ket\W    ,
\label{dSC}
\ee
with the superconductor order parameter
\be
\langle \Psi | d_{ij} c_{i\uparrow} c_{j\downarrow}   |\Psi\rangle  =
\sqrt{Z_b'} e^{i\varphi}\sin\theta\cos\theta.
\label{scop}
\ee

It is worthwhile to reemphasize the following point  which has
important implications in interpreting both experiments and
numerics of  Hubbard like models. The fermions and hole pairs have
charges $+e$ and $+2e$  respectively. Their  number is counted
from the  correlated half filled (RVB)  vacuum. The  operators $f$
and  $b\nd$ should not be confused with  the electron and
Cooper pair operators $c^\dagger_s$ and $c^\dagger_\uparrow
c^\dagger_\downarrow $ respectively, whose numbers  are counted up
from  the electron vacuum. In numerical calculations, it is  preferable to use the operator $b^\dagger$,
as defined by the Hubbard plaquette eigenstates,
as the superconducting order parameter. It should have larger matrix elements  than
the customary d-wave pairing operator $d_{ij} c_{i\uparrow} c_{j\downarrow}$.

\subsection{Underdoped to Overdoped Transition}
\label{ud-od}
Throughout this paper we restrict ourselves to low doping, i.e. a
small number of hole pairs and hole fermions per plaquette.
Nevertheless, the plaquette states basis leads us to expect an
interesting transition at higher hole doping for the following
reason.

When the chemical potential is large enough to bring the hole pair
(2 electron) state to be lower than the 4 electron vacuum, a {\em
vacuum crossing} takes place. On a single plaquette, the
vacuum crossing is when the levels of zero and one boson intersect.
Since hole pairs are somewhat larger in size than a single
plaquette, the lattice vacuum crossing should take place at somewhat
less than $x=0.25$ holes per square lattice site.

Once the two hole state turns into the new vacuum $\ket\W'$, all
excitations are defined with respect to it using different boson  and
fermion  creation operators. For example, the old RVB vacuum
becomes a Cooper pair excitation above the new vacuum.
 \be
 \ket\W\approx (\cP \sum_{ij} d_{ij}c^\dagger_{i\uparrow}  c^\dagger_{i\downarrow}+\ldots) \ket\W'
 \ee

It is plausible that the  vacuum crossing is a true quantum phase
transition, and not merely a mathematical artifact of using
different plaquette bases to construct the same ground state. A candidate for such a phase transition is
the restoration of square lattice
symmetry, if this symmetry is truly broken by plaquettization in the underdoped regime as mentioned in Section \ref{sec:lts}.

The overdoped side is far from half filling, where  effects of
the Gutzwiller projection are small. That is to say, the
eigenstates can be approximated by applying electron operators to
the electron vacuum.  With interplaquette hybridization of the two
electron plaquette vacua,  the ground state can be adiabatically
connected to the quarter filled electron Fermi surface. In the
absence of superconductivity it will exhibit a {\em large}
(Luttinger-theorem obeying) Fermi surface.

\section{Construction of  The Effective Hamiltonian}
Having described the low lying plaquette states, we are faced with the challenge
of constructing an effective  Hamiltonian for the full lattice.
Motivated by the pair binding on a plaquette,  one  might
initially wish to compute the  effective
hopping  of a hole pair  between plaquettes using
second order perturbation theory in the interplaquette hopping  $\tp$.

This naive  approach yields pair hopping of order $J_c\propto
{{\tp^2}\over{\D_b}}$. The perturbative   expansion is  controlled
by $\tp/\D_b$. This suggests failure of perturbation theory for
$\tp=t >>\Delta_{b}$. Indeed, by looking at the exact spectrum of
two connected plaquettes, we find that second order perturbation
fails in a sizable
domain of $t'<t$. {\em Does this imply dissociation of the
local bosonic correlations in the  translationally invariant
lattice?}

{\em We argue no. }

There is convincing evidence from various  numerical approaches, that two holes remain on the same plaquette
in $\sqrt{26}\times\sqrt{26}$ lattices\cite{PB} and on $8\times 6$ t-J ladders\cite{WS}. However, in order to
understand how many pairs behave on the infinite lattice, we must determine  the pair hopping energy,
and derive their effective Hamiltonian.

A suitable approach for this task is provided by
the Contractor Renormalization (CORE) method\cite{core} described below. The small
parameter of CORE is the ratio  of the
hole pair separation, i.e. coherence length  to the range of the effective interactions.

\subsection{Contractor Renormalization Algorithm}
\label{sec:core}

Given a microscopic Hamiltonian $\cH$ on the square lattice we choose
a plaquette covering and proceed by the following steps:

Step 1: {\it Defining the reduced Hilbert space}.
We diagonalize $\cH$ on a single plaquette and truncate all states above
a chosen cutoff energy. This leaves us with the lowest $M$ states $\{\ket{\a}\}_1^M$.
The reduced lattice Hilbert space is spanned by
tensor products of retained plaquette states $\ket{\a_1,\ldots,\a_N}$.
A case in point is the Hubbard model spectrum, which for the
half filled case has 70 states. We truncate 66 states and keep the
ground state and lowest triplet, i.e. $M=4$.
Thus, the Hilbert space is considerably reduced at the first step.

\begin{figure}[htb]
\begin{center}
\vspace*{10pt}
 \leavevmode \epsfxsize0.95\columnwidth
\epsfbox{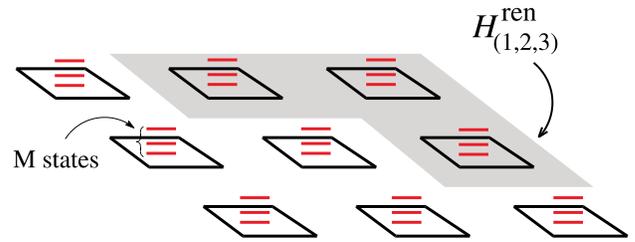}\vskip1.0pc
\caption{{\bf The reduced Hilbert space of a plaquette cluster used in CORE}.
The calculation of $H^{ren}_{1,2,3}$ requires diagonalization of
the Hubbard model on the cluster highlighted by the shaded region.
$H^{ren}$ reproduces the exact spectrum
within the reduced  Hilbert space.
 }
\label{fig:core}
\end{center}
\end{figure}

Step 2: {\it The Renormalized Hamiltonian of a  cluster}. The
reduced Hilbert space on a given connected cluster of $N$
plaquettes is of dimension $\cM=M^N$.  See Fig.~\ref{fig:core} for
an illustration. We diagonalize $\cH$ on the cluster and obtain
the lowest $\cM$ eigenstates and energies: $(\ket{n},\e_n)$,
$n=1,\ldots,\cM$. The wavefunctions $\ket{n}$ are projected on the
reduced Hilbert space and their components in the plaquette basis
$\ket{\a_1,\ldots,\a_N}$ are obtained. The projected states $\psi_n$ are
then Gramm-Schmidt orthonormalized,  starting from the ground
state upward. \be \ket{\tilde\psi_n}={1\over {Z_n}}\left(
\ket{\psi_n}-\sum_{m<n}\ket{\tilde\psi_m}
\braket{\tilde\psi_m}{\psi_n}\right), \label{gs} \ee where $Z_n$ is
the normalization. The renormalized Hamiltonian is defined as
\be
\cH^{ren}\equiv \sum_{n}^{\cM}
\e_n\ket{\tilde\psi_n}\bra{\tilde\psi_n},
\label{Hren}
\ee
which ensures that
it reproduces  the lowest $\cM$ eigenenergies exactly.

Representing $\cH^{ren}$ in the real space plaquette basis
$\ket{\a_1,\ldots,\a_N}$, defines the (reducible) interplaquette
couplings and interactions.

Step 3: {\it Cluster expansion}. We define connected $N$ point
interactions as:
 \be h_{i_1,\ldots,i_N}=H^{ren}_{
\langle{i_1,\ldots,i_N\rangle}}-\sum_{\langle{i_1,\ldots,i_N'\rangle}}
h_{i_1,\ldots,i_N'}, \label{hcon} \ee
where the sum is over connected subclusters of $\langle i_1,\ldots, i_N\rangle$.
The full lattice effective Hamiltonian can be expanded as the sum
\be
\cH_{eff}=\sum_i  h_i + \sum_{\langle ij\rangle}h_{ij}
+\sum_{\langle ijk\rangle}h_{ijk} + ... \label{c-expansion}
\ee

$h_i$ is simply a reduced single plaquette hamiltonian. $h_{ij}$
contains nearest neighbor couplings and corrections to the
on-site terms $h_i$. $h_{ijk}$ contains three site couplings and
so on. $h_{i_1,\ldots,i_N}$ will henceforth be called {\em range-N interaction}.
We expect on physical grounds that for a proper choice of a
truncated basis, range-N interactions will decay rapidly with
$N$. This expectation  needs to be verified on a case by
case basis.

Morningstar and Weinstein, by retaining up to range-3
interactions\cite{core}, demonstrated that the  CORE renormalization group flow,
obtains an excellent value for the ground state
energy of the spin-$\half$ Heisenberg chain.  This is encouraging,
since the spin half chain has long range, power-law decaying spin
correlations. Pieckarewicz and Shepard\cite{PS} tested CORE for
the 12 site spin-$\half$ Heisenberg ladder. They got better than
1\% accuracy for all lowest 64 states using a plaquette basis
keeping only up to range two interactions.

In general, there is no apriori quantitative estimation of the
truncation error. Nevertheless, if it decays rapidly with  interaction range,
we deduce that there is a short {\em coherence length} related to our local degrees of freedom,
e.g. in our case the hole pair bosons and the triplets (bound states of two spinons).


\subsubsection{CORE of Wave Function Correlations }
The CORE process is designed to reproduce the low lying spectrum,
while the wave functions may be significantly distorted during  the
truncation of the Hilbert space. {\em Does this hinder calculation
of   correlation functions within this scheme? } The answer
depends on the operators whose correlations we wish to calculate.

The correlations are calculated by adding external source
terms to the original Hamiltonian,
\be
\cH[\eta]=\cH[0]+\sum_i
\eta_i\hat{O}_i, \ee
where $\hat{O}_i$ is a microscopic linear perturbation
operator whose  correlations we wish to determine.

We apply the  CORE cluster expansion of Eq. (\ref{c-expansion} )  to  the
perturbed Hamiltonian  and obtain $\cH^{eff} [\eta]$   which is  expanded to
linear order in $\eta_i$,   \bea
\cH_{eff}[\eta]&=& \sum_j \left( h_j+\eta_i\hat{O}^{(1)}_{ij}
\right)  + \sum_{\langle
jk\rangle}\left(  h_{jk} +\eta_i\hat{O}^{(2)}_{ijk}
\right)\nonumber\\
&&~~~~+ \sum_{\langle jkl\rangle}\left( h_{jkl} +  \eta_i\hat{O}^{(3)}_{ijkl}
\right) ......+\cO(\eta^2) \nonumber\\ &\equiv &  \cH_{eff}^0 + \sum_i \eta_i
\hat{O}_i^{eff} +\cO(\eta^2)\nonumber\\ \hat{O}_i^{eff} &=& \hat{O}^{(1)}_i +
\sum_j \hat{O}^{(2)}_{ij}+\sum_{jk}  \hat{O}^{(3)}_{ijk}+ \ldots
\label{Oeff}
\eea
 $\hat{O}_i^{eff}$  represents   the linear perturbation $\hat{O}_i$
in the truncated Hilbert space,
 \be
 \hat{O}_i^{eff}\equiv
\sum_{n,m}^{\cM} \langle n |\hat{O}_i |m\rangle
\ket{\tilde\psi_n}\bra{\tilde\psi_m},
\ee
The two-point dynamical correlation function  at low
temperature $T<<\epsilon_{max}$ is given by the Lehmann representation
\bea
S_{ij}(\omega) &\equiv &\nonumber\\
{2\pi\over Z  N} &&\sum_{nm}
e^{-{\epsilon_n\over T}}  \langle n | \hat{O}_i |m\rangle \langle m| \hat{O}_j
|n\rangle  \delta(\omega +\epsilon_n-\epsilon_m )\nonumber\\
\simeq   {2\pi\over Z N} &&\sum_{nm}^{trunc}  e^{-{\epsilon_n\over T}}
\langle  \tilde \psi_n| \hat{O}^{eff}_i |\tilde \psi_m \rangle
\langle  \tilde \psi_m| \hat{O}^{eff}_j|\tilde \psi_n \rangle \nonumber\\
&&~~~~~~~~~\times \delta(\omega
+\epsilon_n-\epsilon_m )
\label{Scorr}
\eea
where $Z$ is the partition function, and the second sum is evaluated in the
truncated Hilbert space, using the cluster expansion (\ref{Oeff}) for the
matrix elements.

Thus   for  the full lattice cluster,  $\hat{O}_i^{eff}$
recovers the exact correlations of the {\em true } low energy eigenstates.
By (\ref{Oeff}) the linearized
cluster expansion   involves  multi-site operators
\be
\hat{O}^{(n)}_{i, i_1,\ldots \i_n}= \partial h_{ i_1,\ldots \i_n}/ \partial
\eta_i
\ee
 A  rapid  decay of  $h_{
i_1,\ldots \i_n}$  beyond a short
truncation range   is essential for the feasibility of the CORE
scheme for the spectrum. Similarly, to calculate the correlations of
$\hat{O}$, we  require  a rapid decay of
$\hat{O}^{(n)}$ with $n$,  which would allow us to  calculate
$\hat{O}_i^{eff}$ by small clusters diagonalizations. We have previously
argued that a small truncation error  results from   a short   coherence
length $\xi  $, for the coarse grained degrees of freedom. For example,
the size of the hole pairs in the slightly doped Hubbard model.  Thus, the
operator $\hat{O}$ should be chosen to have  large matrix elements within the
reduced Hilbert space,  and  the multi-site operators
$\hat{O}^{(n)}_{i_1,\ldots \i_n}$  should decay rapidly  beyond the
range of $\xi$.

In other words,   {\em if  the truncated  wave functions retain the relevant
local operator content,  the cluster expansion for
the operators   converges rapidly, and the effective Hamiltonian can reproduce
the long wavelngth correlations correctly.}  In this paper, the
truncated plaquette states, for example, contain d-wave hole pairs on
plaquettes.  If these hole pairs  turn out to be  tightly bound in the exact
eigenstates of the full lattice,  their  creation operator  has small
multi-site corrections  in the renormalized basis,  i.e. it has a  rapidly
decaying cluster expansion.  In this case, long range d-wave pair correlations
are well represented (up to an onsite renormalization factor),
by   the  boson-boson  correlations of the  Four Bosons
model.

\subsection{Lattice Translational Symmetry}
\label{sec:lts}
The CORE algorithm formally requires explicit breaking of  lattice
translational symmetry at the first step.
 The   plaquette lattice vacuum  breaks lattice
 translational symmetry as follows: each plaquette vacuum contains a triplet
pair contribution, but the product state does not contain
interplaquette triplets, and hence differs from the state
translated by one square lattice spacing. In order to restore the
lattice symmetry, interplaquette triplet pair correlations can be
reintroduced by  triplet pair creation operators in the effective
Hamiltonian.

\begin{figure}[htb]
\begin{center}
\vspace*{13pt}
 \leavevmode \epsfxsize0.75\columnwidth
\epsfbox{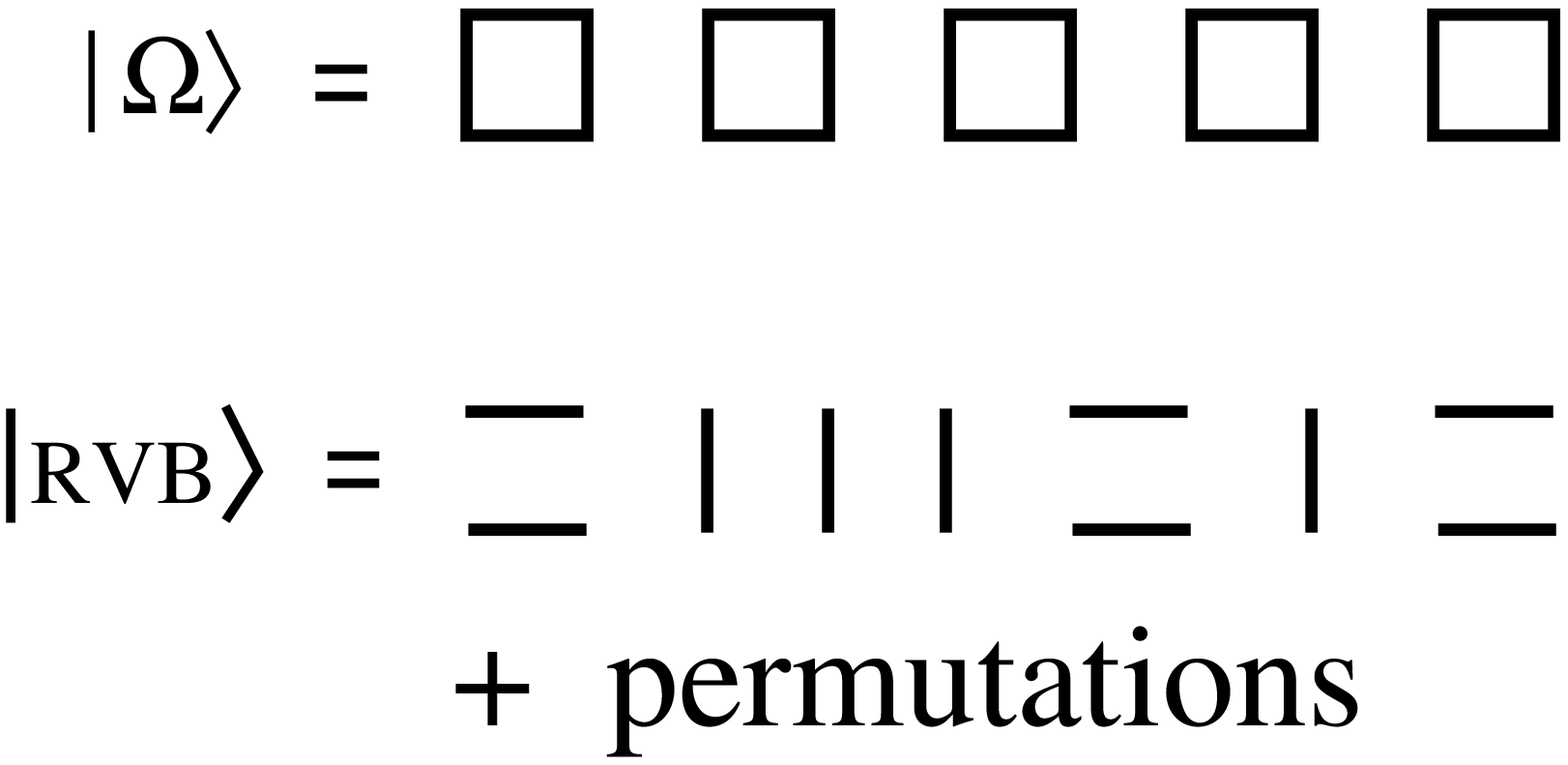}\vskip1.0pc \caption{{\bf Restoration of
translational symmetry on the ladder}.   $\ket\W$ is the plaquette
lattice vacuum  which breaks two fold lattice translational
symmetry.  $|\mbox{RVB}\rangle$ is the dimer Resonating Valence
Bonds state which has translational symmetry, and is related to
$\ket\W$  by an exponential of triplet pairs, see
Eq.(\protect{\ref{RVB}}).
 }
\label{fig:rvb}
\end{center}
\end{figure}

As an illustration, let us consider the plaquette
vacuum of  the two leg ladder, in Fig.\ref{fig:rvb}  which can be
written in the form
 \be
 \ket\W= \cP
  \exp \left( {1\over 3} \sum_{i} t^\dagger_{2i\alpha}  t^\dagger_{2i+1,\alpha}\right)
  \prod_i |0\rangle_i,
  \ee
  where  $|0\rangle_i$ is a  singlet  on rung $i$, and $t^\dagger_{i\alpha}|0\rangle_i$ is
 a rung triplet. $\cP$ projects out multiple occupation of triplets.

The translational invariant RVB state in Fig.\ref{fig:rvb}, can be
constructed from $\ket\W$ by applying the operator
 \be
 |\mbox{RVB}\rangle=
 \cP \exp \left( {1\over 3} \sum_{i} t^\dagger_{2i-1\alpha}  t^\dagger_{2i,\alpha}\right) \ket\W.
 \label{RVB}
 \ee
 In the triplet bosons representation of the Heisenberg exchange there are anomalous
 inter-plaquette terms $t_i^\dagger t_i^\dagger+ t_i t_j$.  In the  mean field
 theory the   Bogoliubov transformation introduces
 a partial ``symmetry restoring'' exponential operator, which resembles
 (\ref{RVB}).  However symmetry cannot be fully restored to
 the wavefunctions in the reduced Hilbert space because of the elimination
 of  higher spin states.

 {\em Is this a problem?}

 Well, it depends on what one is interested in.
 CORE is constructed to obtain accurate effective interactions.
 Unphysical symmetry breaking effects can be introduced
by  truncating the longer range interactions, as explained by a toy model
in Appendix \ref{app:tightbinding}.
Therefore  it is hard
 to {\em rule out} a physical ``plaquettization'' of the true ground state.  Incidentally,  such
 a fourfold discrete symmetry
 breaking is consistent with Berry phase arguments\cite{Haldane} for
 the spin liquid
 phase of spin half Heisenberg models.

 The symmetry breaking, appears as
 minigaps near
 the edges  of the plaquette lattice Brillouin  zone (PLBZ)
 $\kappa_x,\kappa_y \in ( -\pi/2, \pi/2)$.
 In Appendix \ref{app:tightbinding}, we see how the minigaps of the tight binding model,
decrease as longer range interactions are included.
For the triplet and hole pair  bosons, minigaps do not matter much since
their low energy states are  around  $(0,0)$ and $(\pi,\pi)$ respectively;
the farthest possible from the PLBZ edges.

On the other hand,  low energy  fermions happen to be  centered
around the PLBZ corner $(\pm\pi/2,\pm\pi/2)$, where the effects of plaquette symmetry
breaking on the spectrum are large. Although by rotational
symmetry, the  two bands which contain the $(\pi,0)$ and $(0,\pi)$ states
are degenerate at the PBLZ corner, the other two bands have minigaps. These would
distort the elliptical shape of the Fermi pockets, an effect which if it exists, could
be detected by angular resolved photoemmission.

\section{The Plaquette Boson-Fermion Model}
We first start with the bosons, and compute their interplaquette
couplings and interactions using CORE. Later we introduce the hole
fermions, whose parameters are taken from published numerical data
on large  clusters, and estimate their coupling to the bosons
using symmetry arguments. Finally we discuss the properties of the
combined Hamiltonian.

\subsection{Computing Boson Interactions}
\label{sec:4BM}
For the purpose of this paper, we have limited the CORE calculations to
range-2 boson interactions, while projecting out the fermion states.  This required
a  modest numerical diagonalization effort of the Hubbard model on up to 8 site clusters.
The resulting range-2  Four  Boson  model can be separated into
bilinear  and quartic (interaction) terms:
\be
 \cH^{4b}=\cH^b[b] +\cH^t[t]+\cH^{int}  [b,t]
 \label{H4B}
 \ee
where the bosons obey local hard core constraints
 \be
b^\dagger_i b_i +\sum_\alpha t^\dagger_{\alpha i} t_{\alpha i} \le 1
\ee
The bilinear energy terms are
 \bea
 \cH^{b} &=& (\epsilon_b - 2\mu) \sum_i b^\dagger_{i}
b^\nd_{i} - J_b \sum_{\langle ij\rangle} \left(
b^\dagger_{i} b^\nd_{j}+\mbox{H.c.}\right) \nonumber\\
\cH^{t} &=&  \epsilon_t \sum_{i\alpha  } t^\dagger_{ \alpha i}
t^\nd_{ \alpha i} - {J_t\over 2} \sum_{\alpha \langle ij\rangle}
(t^\dagger_{\alpha i}t^\nd_{\alpha j} + \mbox{H.c.})  \nonumber\\
&&~~- {J_{tt} \over 2} \sum_{\alpha \langle ij\rangle}
(t^\dagger_{\alpha i}t^\dagger_{\alpha j} + \mbox{H.c.}).
  \label{4BM}
\eea

\begin{figure}[htb]
\begin{center}
\vspace*{13pt}
 \leavevmode \epsfxsize1.0\columnwidth
\epsfbox{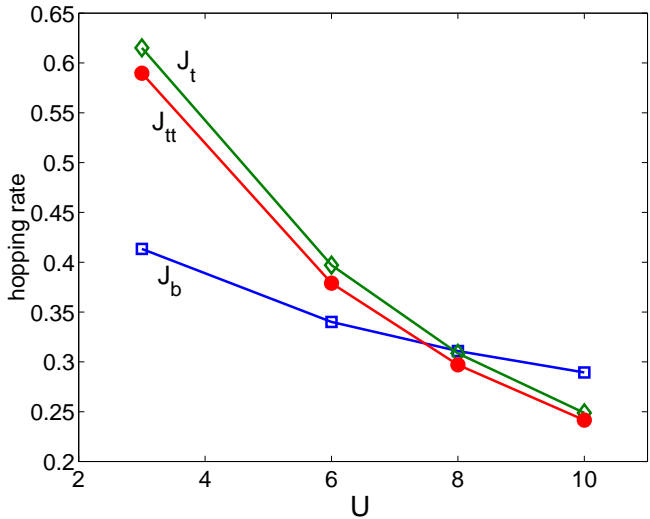}\vskip0.6pc \caption{{\bf Boson hopping energies versus Hubbard $U$.}
$J_t$, and $J_{tt}$ are the magnon's normal and anomalous hopping energies. $J_b$ is the hole
pair hopping energy.
The intersection region near $U=8$ is close to the projected SO(5) symmetry point.
All energies are in units of $t$.
}
\label{fig:pars}
\end{center}
\end{figure}

In Fig.~\ref{fig:pars} we compare the magnitudes of the magnon hoppings $J_t,J_{tt}$ and the
hole pair hopping $J_b$
for a range of $U/t$. First, we observe that $J_{t}\approx J_{tt}\approx 0.6 J$, i.e.
the magnon terms have similar form as those previously obtained for the  Heisenberg model in the
bond operator\cite{gopalan}, and  plaquette operator\cite{Shepard}
representations. Second, the region of intersection  near $U/t=8$,
is close to the {\em projected SO(5)symmetry
point}. We emphasize that
although there is {\em no quantum SO(5) symmetry} in  $H^{4b}$, there is
an approximate equality of the bosons hopping energy scales.
This equality which was  assumed in the
pSO(5) theory\cite{pSO5}, previously appealed
to phenomenological considerations. Here, the equality emerges
in a physically interesting regime of the Hubbard model and has important consequences on
the phase diagram as shown below.

$\cH^{int}$ includes nearest neighbor triplet-triplet, pair-pair,
and pair-triplet interactions. In Appendix
\ref{app:4BM}, $\cH^{4b}$ with all its terms is displayed in its full glory,  and
a table of its computed coupling constants is provided
for the  values of $U/t=3,6,8,10$.
We also compute  the truncation error of discarding
range-3 terms. This is done by comparing the Hubbard model at $U/t=6$, with 0 and 2 holes on 12 sites,
to that of corresponding range-2 $\cH^{4b}$.
Relative shifts of  less than $1\%$  in
the ground state and first excitation energies, correspond to a very small truncation error.

\subsection{Mechanism of Superconductivity}
\label{sec:mechanism}
 There are two important effects which together can
lead to supercondutivity: (i) Pairing, and (ii) Bose condenstion
of the pairs.  An important energy scale for both effects is the
pair hopping rate $J_b$.

The small range-3 truncation error was found
at a large  interaction  $U/t=6$,  where there is actually no pair binding on a
single plaquette (see Fig.~\ref{fig:pairbinding}).  The convergence of effective interactions,
implies   {\em short boson coherence lengths} $\xi_t,\xi_b$.   $\xi_t$ is
the distance between
spinons  (localized spin half configurations) which comprise a magnon.   $\xi_b$
is the hole pairing distance. Both coherence lengths
appear  to be of the order of one plaquette size.
This conclusion is supported by  numerical observation of short distance (lattice constant)
correlations between two holes on large  lattices\cite{PB,WS}.
It is interesting that the short pair  coherence length
is dynamically generated in the Hubbard model, even for  $U/t>4.5 $
where the pair binding energy on an isolated plaquette is positive.

{\em Why is $\xi$ so short?}

There are two effects which bind
pairs: a classical magnetic energy  from minimizing the number of broken
Heisenberg bonds, and a quantum kinematic pairing for holes moving
on two sublattices of a quantum disordered antiferromagnet. The
first effect is supported by finding pair binding on a single
plaquette. However, this energy also favors clumping many holes together.
The quantum pairing effect was proposed by Weigmann, Lee and
Wen, who integrated out spin fluctuations in a quantum disordered phase, to induce a
long range electrodynamical attraction between
holes on opposite sublattices\cite{WLW,AL}.  The kinematic effect produces pairing rather
than phase separation, and is robust
against additional short range repulsion. It also can explain pair
binding on large clusters in a  regime of $U/t>4.5$\cite{pairbinding}.

{\em Bose condensation.}

The relative large hopping in  the pair kinetic energy $- J_b\sum_{\langle i,j\rangle}
b^\dagger_i b_j$ is crucial for understanding the  cuprate phase
diagram.

(i) The pair kinetic energy  competes effectively with the
antiferromagnetic order. While uncorrelated  single fermion
kinetic energy is not inhibited by the presence of long range antiferromagnetic order
(in fact it strengthens it by a Nagaoka-like mechanism), the pair
kinetic energy is substantially lower in a background of short range
singlet correlations. This effect was clearly demonstrated in variational
Monte Carlo studies of pair kinetic energy in doped RVB wavefunctions\cite{havilio}, and
is  also a property of the variational treatment of the four boson model.

The destruction of antiferromagnetic order into a quantum spin liquid with massive triplets,
also helps in the kinematical pairing process as discussed above.

(ii) Having destroyed antiferromagnetic order,
the pair kinetic energy competes with charge localization due to
disorder,  or solidification (charge density wave), and with disintegration into
unbound hole fermions.

(iii)  A large $J_b$   stabilizes a superconducting phase at finite temperatures.
It  determines  the superfluid density $\rho_s = 2 J_c
|\langle b\rangle|^2$, and the phase ordering
transition temperature\cite{EmeryKivelson}   $T_c \approx \rho_s $.

\subsection{Four Boson  Mean Field Theory}
The mean field theory  is separated into two parts: (i)  Calculation of
the order parameters as  a function of doping, using variational coherent states. (ii)
Determination of magnon resonance energy from a soft interaction version. The results are qualitatively similar
to the projected SO(5) phase diagram\cite{pSO5,pSO5-calcutta}.

\begin{figure}[htb]
\begin{center}
\vspace*{13pt}
 \leavevmode \epsfxsize1.0\columnwidth
\epsfbox{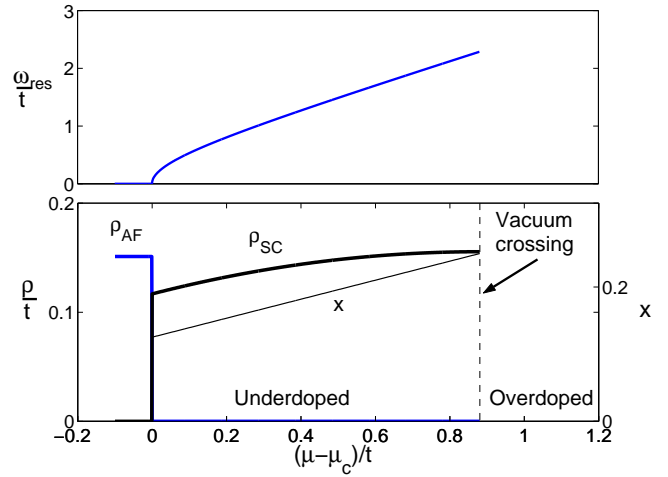}\vskip0.6pc \caption{{\bf Variational Solution of the Four Boson  model.}
Results correspond to  Hubbard interaction strength $U/t=8$.
$\omega_{res}$ is the antiferomagnetic  resonance energy, $\rho_{AF}$  is the spin stiffness
in the antiferromagnetic phase,
and $\rho_{SC}$ is the superfluid density in the superconducting phase.
$\mu-\mu_c$ is the chemical potential difference from the first order transition at $\mu_c$,
and $x$ is the hole density. The estimated vacuum crossing point is discussed in Section
\protect{\ref{ud-od}}.
}
\label{fig:mfbosU8}
\end{center}
\end{figure}

Here,  we choose $U=8t$, and evaluate the energy of the  full
boson Hamiltonian (\ref{4b-hopping},\ref{4b-int}) in the variational coherent states   $\psi^{afm}(\theta)$
and $\psi^{d-sc}(\theta)$ of Eqs. (\ref{Neel}) and (\ref{dSC}) respectively. These states
represent the antiferromagnetic and superconducting phase. The critical chemical potential $\mu_{c}$, where
the ground state energies cross,
and  quantum fluctuation
angle $\theta(\mu)$ are determined by  minimizing the energy.
The spin stiffness and superfluid density are given respectively by
\bea
\rho_{AF}&=& 2J_t \langle t\rangle^2,\nonumber\\
\rho_{SC}&=& 2J_b \langle b\rangle^2,
\eea
where we use Eqs.(\ref{stagm},\ref{scop}) for
the magnon and hole pair expectation values.
These coefficients, which determine the transition temperatures, as well as the doping concentration
$x$ are
plotted as a function of chemical potential
in Fig.\ref{fig:mfbosU8}. We emphasize that the results should not be quantitatively
compared to experiment, since they are  variational approximations to a simple model,
and neglect effects of low energy hole fermions.

The variational theory  yields a first order transition between zero doping and $x_{c}=x(\mu_c)
\approx
0.125$, where
the staggered magnetization abruptly vanishes and the superfluid density jumps to a finite value.

For charged holes, this first order transition (phase separation),
is forbidden by long range Coulomb interactions.
Instead one expects
high compressibility, incommensurate mixed phases and stripes\cite{EmeryKivelsonZachar}
in the intermediate doping regime $x\in (0,x_c)$.

Even a weak disorder potential is very efficient in breaking the intermediate phase
into ``quantum melts''\cite{SAK,Kapitulnik}, i.e.  puddles of
superconductor inside antiferromagnetic domains.

Above $x_{c}$, the superfluid density
increases  with doping, in agreement with London  penetration depth measure\~ments\cite{Uemura},
The overdoped regime is beyond the
expected vacuum crossing point (see Section \ref{ud-od}).

The magnon dispersion in the superconducting phase is obtained by decoupling
a soft core interaction\cite{pSO5,pSO5-calcutta},
\be
\cH^{int} = W  \sum_i  : \left( b^\dagger_i b^\nd_i + \sum_\alpha t^\dagger_{\alpha i}
 t^\nd_{\alpha i} \right):~~,
 \ee
 where $W$ is fitted to yield the order parameter magnitudes calculated variationally.

In the superconductor,
 the magnons acquire a gap at the antiferromagnetic resonance $\omega_{res}$
 which  increases  with doping as
 \be
  \omega_{res}= 2\sqrt{(\mu-\mu_c)(\mu-\mu_c+ 2J_t)}\propto \sqrt{x-x_c}.
  \ee
  This dependence, as plotted in Fig.\ref{fig:mfbosU8},  is  qualitatively consistent  with
  inelastic neutron scattering data\cite{afm-res}.

\subsection{Fermion  Hamiltonian}
\label{subsec:fer-ham}
 In the previous  section we have computed the bosonic interactions of
(\ref{H4B})  from the Hubbard model using CORE.  In that
computation,  we have eliminated the fermion (single hole)
states. we expect however that for the two dimensional
square lattice, low energy fermion excitations are important.
While the fermion holes short range effects on the boson couplings
were included in the range-2 CORE calculations,
their long wavelength  excitations, require
diagonalizing larger clusters  which are beyond
this paper's computational scope.
We therefore resort to including the hole fermions dispersion
``by hand'' i.e. use the single hole band structure computed previously for large clusters.
We then  estimate their interactions with the bosons.

It is important  to emphasize that the definition of the
hole pair bosons and  the hole fermions is simply a
matter of separation: two hole fermions are on {\em different}
plaquettes. When they hop into the same
plaquette they turn into one boson via the Andreev coupling defined below.

\begin{figure}[htb]
\begin{center}
\vspace*{13pt}
 \leavevmode \epsfxsize0.95\columnwidth
\epsfbox{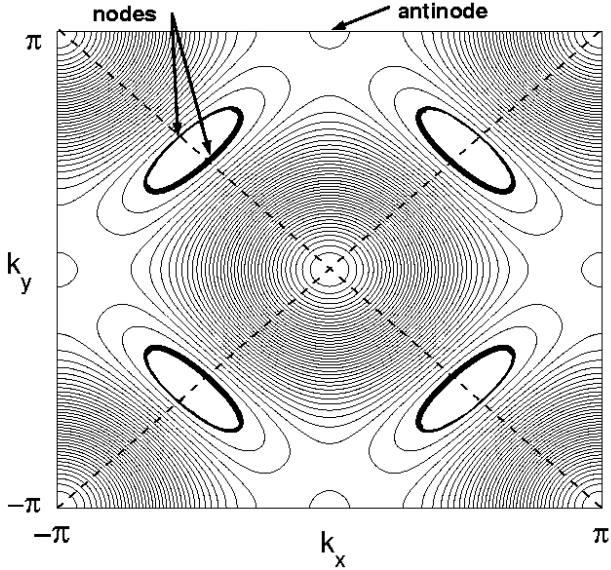}\vskip0.6pc \caption{{\bf Hole fermions
bandstructure}.   Contours of the single hole spectrum
Eq.(\protect\ref{Hf} ) modelled by fitting to published numerical
results. For a dilute number of free holes,
Fermi pockets will be created around the points $(\pm \pi/2,\pm\pi/2)$. A flat valley
 near the magnetic zone edge dominates the low energy spectrum.
 }
\label{fig:band}
\end{center}
\end{figure}

For the relevant range of $U/t$ the numerically determined band structures
for the single hole can be
fit by two hopping energies
 \bea
\cH^{f} &=&  \sum_{\bk s}  ( \epsilon^f_\bk-\mu) f^\dagger_{\bk s}  f_{\bk s}
, \nonumber\\
\epsilon^f_\bk&=&  t'  (\cos(k_x a) +\cos(k_y a) )^2 \nonumber\\
&&~~~ +t'' (\cos(k_x a) -\cos(k_ya) )^2   .
 \label{Hf}
 \eea
$\bk$ runs over the square lattice Brillouin zone. See Fig.~\ref{fig:band}. The values $t'\approx J$, and $t''\approx 0.1J$
are taken from the numerical
Quantum Monte Carlo data for the t-J model on a $24\times
24$ lattice\cite{tJ24}, find that for the physically relevant range of $J\in [0.4t,0.6t]$,
the dispersion values are $t'\simeq 0.7J$, $t''\simeq ~0.1t'$.

The magnitude of $t'\approx J$ (rather than the bare value $t$)
and the position of the  minima on the magnetic zone edge  $
(\pi,0)-(0,\pi)$  were explained by theories of holes in the short
range antiferromagnetic environment\cite{hole-theory}. The
semiclassical theory\cite{AL}  finds that holes are highly
dressed local spin polarons, which effectively hop
on one sublattice.

From the CORE's perspective, the flat valley  between $ (\pi,0)$ and
$(0,\pi)$ is related to  the original degeneracy between the two
lowest plaquette fermions. Thus we expect the wavefunctions of the
fermions on the lattice to contain a large component of these two
states. Consequences of this on the quasiparticle weight and
possible staggered orbital currents were mentioned in Section
\ref{SHF}.

The holes have hard core interactions among themselves, and with the bosons.
At low doping however it is still meaningful to describe their
states by excitations about small Fermi
pockets around $(\pm\pi/2,\pm\pi/2)$.

\begin{figure}[htb]
\begin{center}
\vspace*{13pt}
 \leavevmode \epsfxsize0.95\columnwidth
\epsfbox{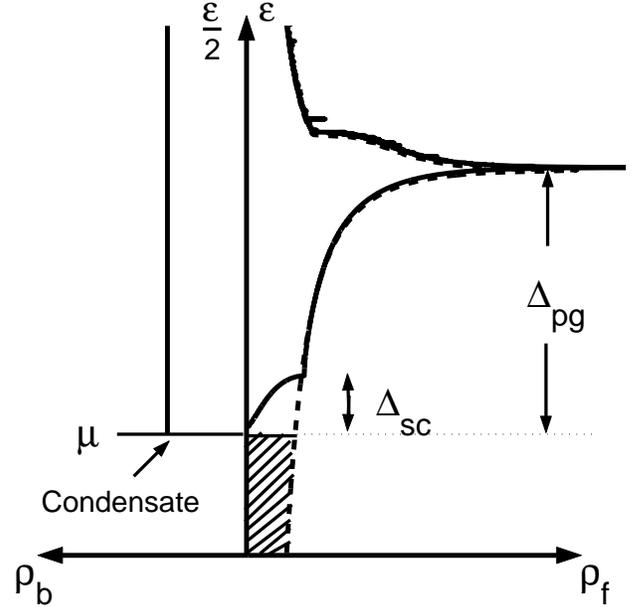}\vskip0.6pc \caption{{\bf Hole fermions and hole pair bosons  density
of states}.   The fermion density of states is calculated for
dispersions Eq. (\protect\ref{Hf})  for the normal state (dashed
line), and superconducting state Eq.(\protect\ref{SC}) (solid
line).   The low energy scale $t''$ creates a large peak in
the hole density of states. Hole pair bosons single particle density of states
is approximated as a constant corresponding to the non interacting
bilinear terms of $H^b$ in Eq. (\protect\ref{4BM}). $\Delta_{sc}$ and $\Delta_{pg}$
refer to the superconducting gap of eq. (38) and pseudogap of eq. (37)
respectively.}
\label{fig:dos}
\end{center}
\end{figure}

The fermion density of states of (\ref{Hf}) is plotted in Fig.~\ref{fig:dos}. We see a large peak at low energies (of order
$4t'' << 4t'$) from  the saddlepoints near  the antinodal points.
These dominate the hole spectral function, and tunneling density
of states at the {\em ``pseudogap''} $\Delta_{pg} $ energy above the
chemical potential. Within this framework, $\Delta_{pg}$ does not
describe the pairing correlation per se. (It only feels the change
in boson density through changes in the common chemical
potential). Even in the superconducting phase where  hole pairs
Bose condense,  near antinodal points  the Bogoliubov
particle-hole admixture is small, and  quasiparticles have a
character of {\em holes} in the RVB vaccuum. This has important
experimental implications:

\begin{enumerate}
\item  {\em Angular Resolved Photoemmission}. The
 large  Fermi surface of electrons, given by  Hartree-Fock  approximations, includes mostly the first
magnetic Brillouin  zone (the diamond connecting antinodal
points). Luttinger's theorem for a Fermi liquid of electrons {\em
excludes} any hole spectral weight outside this area. In contrast,
spectral weight of our fermions can be found anywhere outside the
small Fermi pockets near $(\pm\pi/2,\pm\pi/2)$. Indeed,  broad
quasiparticle weight, above $T_c$  has been observed in
photoemmission data at momenta half way on the line $(\pi,0) \to
(\pi,\pi)$\cite{pi0-pipi}.

A direct evidence of small Fermi pockets would be sharp gapless
quasiparticle modes on both sides of $(\pi/2 ,\pi/2 )$,. The
``shadow'' quasiparticles closer to $(\pi,\pi)$ are harder to
observe than the ones closer to $(0,0)$, because of vanishing
quasiparticle weight as discussed following Eq.(\ref{ZQ2}).

\item {\em Tunneling conductance} should exhibit an inherent asymmetry between
injecting electrons (positive bias) and injecting holes (negative
bias). The negative bias peak at the pseudogap voltage is larger
than the positive peak, since injecting electrons is suppressed by
Hubbard interactions. In other words, electrons can only be
injected into existing holes, whose density is of order $x$, at
low doping the ratio of weights should scale with $x$. A  review
of (unsymmetrized)  tunneling data published by several
groups\cite{tunneling,Kapitulnik-PC} reveals such an asymmetry,
although we have not seen yet a systematic study of its doping
dependence in the literature.
\end{enumerate}

\subsection{Boson-Fermion Couplings}
Couplings between bosons and fermions can be derived by microscopic considerations  and
symmetry.

\begin{figure}[htb]
\begin{center}
\vspace*{13pt}
 \leavevmode \epsfxsize0.75\columnwidth
\epsfbox{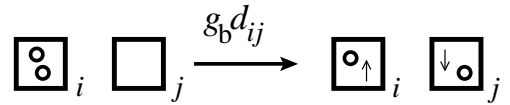}\vskip0.6pc \caption{{\bf Andreev coupling
between hole pairs and fermions}.   The microscopic origin of the
hole pair-hole fermion coupling is a simple unbinding process. Because
of the hole pair d-wave symmetry, the coupling matrix element $d_{ij}$ is
odd under $\pi/2$ rotations on the lattice.
 }
\label{fig:andreev}
\end{center}
\end{figure}

Taking into account the $d$-wave
symmetry of  the hole pair state yields an Andreev coupling (see Fig.\ref{fig:andreev}):
\be
\cH^{bf} = g_b\sum_{\bk,\bq }\left(  d_{\bk+\bq/2}  b^\dagger_{\bq} f^\nd_{\bk\uparrow}
f^\nd_{-\bk+\bq\downarrow}+\mbox{H.c}\right),  \label{Hbf}
\ee
 where $d_\bk=\cos(k_x) -\cos(k_y) $, and
 $b^\dagger_\bq=\sum_i^{plaq} e^{i \bq \bx_i} b^\dagger_i$
 is  a Fourier component on the plaquette
 lattice.

In the superconducting phase $\langle b\rangle\ne 0$.  This
implies a proximity induced pairing of fermions in the small
pockets,  and  an opening of a superconducting gap with the
Bogoliubov dispersion
 \bea
 E_\bk &=&   \pm \sqrt{ (\epsilon_\bk-\mu)^2 + \Delta_\bk^2 }\nonumber\\
  \Delta^{sc}_\bk &=&  g_b d_\bk \langle b\rangle.
 \label{SC}
 \eea

 To be consistent with  the range-2 CORE method, we must
 not include close-by holes on nearest neighbor plaquettes. These excitations were
 already taken into account in the effective hole pairs hopping
 energy. The remainder Andreev coupling is therefore between
 second nearest neighbor hole fermions, with a coupling constant $g_b \le 0.1
 J_b$, estimated from the magnitude of the range-3 terms (see Appendix \ref{app:4BM}).

 We emphasize that $\Delta_{sc}$ is not the``usual''
BCS gap, since it couples to hole fermions, not electrons. Through
its dependence on the Bose condensate order parameter $\langle
b\rangle_{Tx} $, we can deduce the transverse quasiparticle
velocity at the nodes $v_\perp = \partial \Delta_\bk /\partial
k_\perp$. $v_\perp\to 0$ at $T_c$, and  it should vary with doping
as
 \be v_\perp \propto
\sqrt{T_c}\propto \sqrt{x} .\ee

At higher temperatures than $T_c$, $\Delta_\bk$  vanishes and a
broadened signature of the small Fermi surface
emerges in the spectral function. In contrast, near antinodal
points,  Bogoliubov particle-hole mixing is negligble  and
spectral weight is due to hole fermions. This is consistent with
photoemmission data which finds that above $T_c$ the gap closes
only in a small region around the nodal
direction\cite{norman-nature}.

The Andreev coupling (\ref{Hbf}) couples the superconducting phase
fluctuations to nodal quasiparticles. Similar interactions were
used to calculate the temperature dependent London penetration
length\cite{mohit}. That calculation found the fermions to
be more dominant at low temperatures than thermal phase
fluctuations in destroying the superfluid density. The effects of
this term on the fermions above $T_c$, were recently argued to
 give rise to marginal Fermi liquid spectral peaks\cite{FT}.

Lastly,  the fermion-magnon coupling is given by
 \be
\cH^{tf} = g_t\sum_{m s \bk \bq }\left( (t^\dagger_{m \bq} +
t^\nd_{-m -\bq}) f^\dagger_{\bk s } f^\nd_{\bk+\bq+\vec{\pi} s+m
}+\mbox{H.c}\right). \label{Htf}
 \ee
 This singlet interaction term, flips fermion spins and scatters them with momentum $(\pi,\pi)$
 while emitting or absorbing
magnons. It produces signatures of the antiferromagnetic resonance
in the fermions self energy\cite{eschrig-norman,pi0-pipi}. (\ref{Htf}) is similar to
fermion-magnon terms which were considered for predicting
antiferromagnetic resonance signatures in tunneling and
photoemmission\cite{abanov-chubukov}.

\subsection{Boson-Fermion Thermodynamics}
The end result of the previous sections is a system of four bosons
and a gas of  hole fermions  in thermochemical equilibrium, i.e.
the charged bosons and fermions share a common chemical potential
$\mu$.  Combining (\ref{H4B}),(\ref{Hf}),(\ref{Hbf}),(\ref{Htf})
yields the complete Plaquette Boson-Fermion Hamiltonian:
\be
 \cH^{PBFM} =
\cH^{4b}[2\mu]    + \cH^{f}[\mu]  +  \cH^{b f} +    \cH^{t f}.
\label{Hpbfm} \ee

In a uniform phase, the fermions and
 and hole pair  bosons  obey a global charge density constraint
 \be 2 n_b (2\mu,T) + n_f(\mu,T)  = x.
\label{constraint} \ee
An important missing parameter, in the absence of a consistent
calculation of the fermions bands, is the relative position  of
the lowest fermion  and hole pair  energies.

Numerical  evidence for $4\times 4$ Hubbard clusters\cite{dag}
show  that for up to three hole pairs, there is a negative pair
binding energy, i.e. the lowest fermion state at $(\pi/2,\pi/2)$
is still above the boson condensate.   However, at finite doping where superconductivity
wins over antiferromagnetism, the repulsively interacting bosons may have higher energy than
the bottom of the fermion bands. This will produce
gapless nodal fermions in the superconductor. Here we shall assume that already at very low
doping these energies match, and  bosons and fermions coexist.

The boson and fermion compressibilities are
 \bea \kappa_b&=&
\partial n_b/ \partial (2\mu),\nonumber\\
 \kappa_f&=&\partial n_f/\partial \mu .\label{kappa}
 \eea
Where $n_b,n_f$ are boson and fermion densities per square lattice site.
The zero temperature fermion compressibility, up to a Landau
parameter correction, is approximately equal to the
Fermi pockets density of states, by (\ref{Hf}):
 \be \kappa_f \sim {1\over {\pi\sqrt{t't''}}} \ee
 The boson compressibility, (using the xy model representation of hard core bosons) is approximately
\be
  \kappa_b \sim {1\over 32  J_b}.
  \label{kb-est}
  \ee
At zero temperature, ignoring boson-fermion interactions,
we use Eqs.(\ref{kappa}) and (\ref{constraint})
to obtain the change in chemical potential to linear order in
doping $x$
\be
\mu(x)-\mu(0) =  {x \over (2\kappa_b + \kappa_f)}.
 \ee

\begin{figure}[htb]
\begin{center}
\vspace*{13pt}
 \leavevmode \epsfxsize1.0\columnwidth
\epsfbox{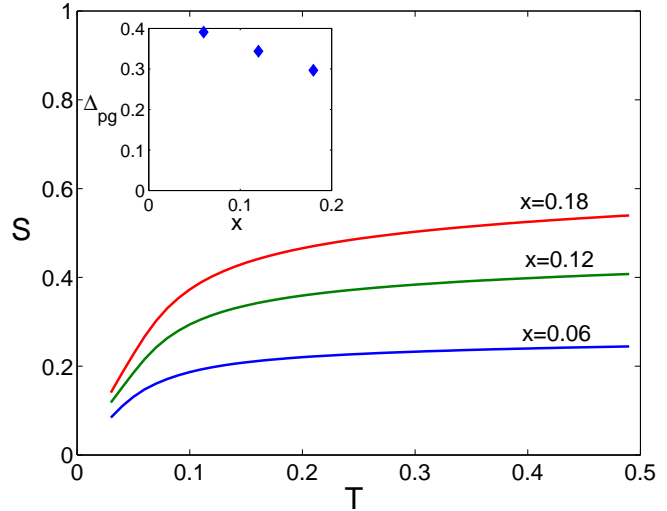}\vskip0.6pc \caption{{\bf Thermodynamics of
the PBFM}.  Excess normal state hole entropy as a
function of temperature for different doping levels,
calculated using Eqs. (\protect{\ref{entropy}}). Inset: The pseudogap energy as a function of
doping. Energies and temperatures are in units of the holes hopping parameter $t'$ of Eq. (\ref{Hf}).}
\label{fig:thermo}
\end{center}
\end{figure}

In the underdoped regime, where $x<<1$, the   energy
distance between
    $\mu$ and the  fermion saddlepoints
    $\bk^{sp} \approx  (\pi, 0)$  defines the {\em pseudogap} $\Delta_{pg} $
    as   measured in tunneling and photoemmission (see Fig. \ref{fig:dos}.     Its doping dependence is simply connected to
    the chemical potential shift
\be
     \Delta_{pg}(x) =E_{\bk^{sp} } - \mu(x,T),
\ee
which yields a steady reduction of the pseudogap as a function of doping as plotted in Fig.~\ref{fig:thermo}

In the normal state above $T_c$, we have a theory of two decoupled, non interacting
gases. For the bosons, we use
a constant density of states $\rho_b(\omega)$,  and for the fermions we choose
$\rho_f(\epsilon)$ from the dispersion (\ref{Hf}), (see Fig.~\ref{fig:dos}). In this
simplified theory, $\mu(T,x)$ can be found using   Eq.(\ref{constraint}) and solving
 \be
  2 \int d\omega {\rho_b(\omega)  \over e^{\omega-2\mu\over T} -1 }
  + \int d\epsilon {\rho_f(\epsilon)  \over e^{\epsilon-\mu\over T} +1 } =x.
  \ee
     The grand potential and entropy are given by
  \bea
   \Omega(T) &=&   -T\int d\omega \rho_b(\omega)  \log \left( 1-e^{-(\omega-2\mu(T))\over T} \right)
    \nonumber\\
    && ~~~~~ + T\int d\epsilon  \rho_f (\epsilon) \log\left (e^{-(\epsilon-\mu)\over T} +1 \right) , \nonumber\\
 S(T) &=&  -\partial \Omega / \partial T.
 \label{entropy}
    \eea
In Fig.\ref{fig:thermo} the excess hole entropy $S(T,x)$ of in
the non superconducting state is shown for the  density of states
given in Fig.\ref{fig:dos}.  The picture which emerges is that  above the superconducting
transition temperature,
bosons  {\em evaporate} into the fermions gas.   The
evaporation is driven by the larger density of states of hole fermions than the bosons.
This evaporation also implies a rapid increase in
magnetic susceptibility. Its effects on transport have not yet
been calculated.

\section{Summary and Discussion}
\label{Discussion} This  paper is primarily aimed at demonstrating
the application of CORE to the Hubbard model, which allows us to
extract its
low energy degrees of freedom and derive the Plaquette  Boson-Fermion  model.
The CORE
calculation could be improved by diagonalizing  larger
clusters within contemporary computational capabilities.
A consistent computation of both the boson and fermion parameters would be  useful.
It would permit systematic studies
of extended Hubbard models and the effects of additional interactions.

The d-wave hole pairs  are already present in
the Hubbard model on a single plaquette. Fortunately, due to the short coherence
length and large hopping rate, the pairs maintain their integrity
in the square lattice.

The PBFM, at the simplest level
of approximation, provides a phase diagram which shares the basic
features of underdoped cuprates: the antiferromagnetic Mott
insulator and a d-wave superconductor with nodal hole fermions. In
the superconducting phase, the local spin one magnons are gapped
at the antiferromagnetic resonance energy, and the remaining
gapless excitations consist of a small density of hole pair bosons
and spin half hole fermions.

The PBFM brings us closer to understanding low temperature
correlations of cuprates. It is amenable to  mean field, low density,
and variational approximations which do not lend themselves
directly to the higher energy Hubbard model and its various extensions.

Here, the PBFM was only preliminarily explored.
It would be interesting to study its thermodynamics
and transport properties in more detail.

\section*{Acknowledgements}
We thank
G. Koren, A. Kapitulnik, M. Norman, M. Randeria, S. Sondhi, S.
Sachdev, M. Weinstein and S-C. Zhang for useful discussions. A.A.
acknowledges the hospitality of the Institute for Theoretical
Physics at Santa Barbara, and  Aspen Center for Physics, and
support from Israel Science foundation,  U.S.-Israel Binational
Science Foundation, and the Fund for Promotion of Research at
Technion.
 \appendix
\section{The complete four boson model}\label{app:4BM}
Here we  present the complete four boson model including all interactions generated
by CORE up to
2 plaquette terms. Coupling parameters are listed for square lattice
and ladder geometries. We then estimate the magnitude of the truncated
three plaquette terms.

The four boson model can be separated into a bilinear part
and a quartic part in the bosonic operators:
\be
\cH^{4b}=\cH^b[b] +\cH^t[t]+\cH^{int}  [b,t]
\label{H4b-app}
\ee
where the bosons obey local hard core constraints
\be
b^\dagger_i b_i +\sum_\alpha t^\dagger_{\alpha i} t_{\alpha i} \le 1.
\ee

The kinetic (bilinear) terms as written in section \ref{sec:4BM} are
\bea
 \cH^{b} &=& (\epsilon_b - 2\mu) \sum_i b^\dagger_{i}
b\nd_{i} - J_b \sum_{\langle ij\rangle} \left(
b^\dagger_{i} b\nd_{j}+\mbox{H.c.}\right) ,\nonumber\\
\cH^{t} &=&  \epsilon_t \sum_{i\alpha  } t^\dagger_{ \alpha i}
t\nd_{ \alpha i}
- {J_t\over 2} \sum_{\alpha \langle ij\rangle}
(t^\dagger_{\alpha i}t\nd_{\alpha j} + \mbox{H.c.})  \nonumber\\
&&~~- {J_{tt} \over 2} \sum_{\alpha \langle ij\rangle}
(t^\dagger_{\alpha i}t^\dagger_{\alpha j} + \mbox{H.c.}).
\label{4b-hopping}
\eea
The higher order interaction terms are
\bea
\cH^{int} &=& V_b\sum_{\langle ij\rangle}n_{bi}n_{bj}
+\sum_{\langle ij\rangle}\big[V_0(t_it_j)^\dagger_0(t_it_j)\nd_0 \nonumber\\
&&+V_1(t_it_j)^\dagger_1(t_it_j)\nd_1
+V_2(t_it_j)^\dagger_2(t_it_j)\nd_2\big] \nonumber\\
&&-J_{bt}\sum_{\langle ij\rangle\alpha}
( b^\dagger_i b\nd_j t^\dagger_{\alpha j}t\nd_{\alpha i}+ \mbox{h.c.} )
\nonumber\\
&&+V_{bt}\sum_{\langle ij\rangle\alpha}
( b^\dagger_i b\nd_i t^\dagger_{\alpha j}t\nd_{\alpha j}
+ b^\dagger_j b\nd_j t^\dagger_{\alpha i}t\nd_{\alpha i} ),
\label{4b-int}
\eea

where $(t_i t_j)\yd_S$ creates two triplets on plaquettes $i$
and $j$, which are coupled into total spin $S$.
When $V_0=2 V_1=-2V_2$ the triplet interactions may be written
using spin-1 operators as $V_2\bS_i\cdot\bS_j$.
Similarly, For $J_t=J_{tt}$, which is close to the value given by CORE,
(see table \ref{tab:all-par}), the bilinear
two site triplet terms may be simplified to $J_t \bn_i\cdot\bn_j$,
with $n_\a={1\over\sqrt{2}}(t\yd_\a+t^\nd_\a)$.

The full Hamiltonian \ref{H4b-app} may serve as a starting point
for various approximations or numerical studies. Its parameters
were computed using CORE from the Hubbard model with
$U/t=3,6,8,10$. The parameters are listed in table
\ref{tab:all-par}.

\begin{table}[h]
\begin{tabular}{|c|c|c|c|c|}
    &  $U=3t $  & $U=6t $ & $U=8t$ & $U=10t$  \\ \hline
  $\epsilon_0$ &  -6.613  & -8.332  & -9.865  & -11.549 \\
               & (-6.019) &(-7.983)&(-9.593) & (-11.324)\\
  $\epsilon_t$ & 0.152   & 0.183  & 0.174  &  0.162 \\
               &(0.192)&(0.263) &(0.253) & (0.233)\\
  $\epsilon_b$ & 1.178   & 2.081  & 3.557 & 5.183 \\
               &(0.440) &(3.212) & (4.835)&(6.567)\\
  $J_t$ & 0.615  & 0.397 & 0.309 & 0.249 \\
  $J_{tt}$ & 0.590  & 0.379 & 0.297 & 0.242\\
  $V_0$ & -0.361  & -0.152 &-0.114& -0.099\\
  $V_1$ & -0.203  & -0.117&-0.095 & -0.082\\
  $V_2$ &  0.214 &  0.099& 0.071 & 0.055\\
  $J_b$ & 0.413  & 0.340 & 0.311 & 0.289\\
  $J_{bt}$ & -0.383  & -0.233 & -0.173&-0.134\\
  $V_{bt}$ & -0.133 &  -0.286 &-0.143& -0.191\\
  $V_{bb}$ &  0.884 &  1.061 & 1.145&1.213\\
\end{tabular}
\vskip0.4pc
\caption{ Parameters for the Four Boson model, in units of $t$
on the square lattice and ladder. The parameters were
computed from the Hubbard model using range-2 CORE.  Values for the ladder are given in
parenthesis where they  differ  from the square lattice.}
\label{tab:all-par}
\end{table}

Note that the on-site terms for the ladder geometry (given in parenthesis
in table \ref{tab:all-par})
differ from the square lattice case due to
contributions of 2 plaquette terms $h_{ij}$.
For example let $\e^0_t$ be the bare on site triplet energy from the
single plaquette spectrum and $\delta\e_t$ the correction due to the
inter plaquette interaction as described in section \ref{sec:core}. The
renormalized on-site energy at site $i$ is $\e^0_t+z_i \delta\e_t$
where $z_i$ is the coordination number of site $i$.
The values of $\e^0_t$ and $\delta\e_t$ may be extracted from the table.
For example:
\bea
\delta\e_t&=&(\e^{square}_t-\e^{ladder}_t)/2 \nonumber\\
\e^0_t&=&\e^{square}_t-4\delta\e_t .
\eea

\begin{figure}[h]
\begin{center}
\vspace*{13pt}
 \leavevmode \epsfxsize0.9\columnwidth
\epsfbox{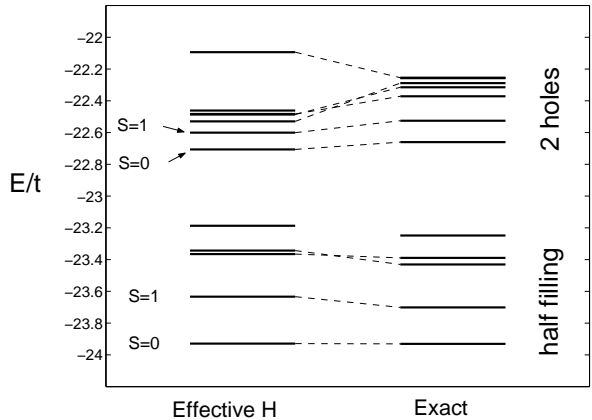}\vskip0.6pc
\caption{{\bf Low energy Spectrum of
exact compared to effective Hamiltonian on 3 plaquettes}. The
comparison is presented for the Hubbard model with U=6t in the 0-hole and 2-hole sectors.
An arbitrary chemical potential was used to
set the 2-hole energies slightly above the plotted 0-hole energies.}
\label{fig:compare}
\end{center}
\end{figure}

{\em Estimation of the truncation error.}
In Fig. \ref{fig:compare} we compare between the low energy spectrum of the
exact and the truncated effective Hamiltonian for 3 collinear plaquettes.
This comparison may be used
to estimate the magnitude of the higher order three plaquette terms $h_{ijk}$
defined by equation (\ref{hcon})
\be
h_{ijk}\equiv H^{ren}_{ijk}-(h_{ij}+h_{jk}+h_i+h_j+h_k)=H^{ren}_{ijk}-H^{eff}_{ijk}.
\ee
Recall that $H^{ren}_{ijk}$ has the exact low lying spectrum of the original Hamiltonian
on the 3 plaquettes. Thus expectation values of $h_{ijk}$ in the ground state
and first excited states are calculated by subtracting energies of $H^{eff}_{ijk}$
from corresponding exact energies of the three plaquette problem.
We estimate $\langle h_{ij}\rangle$ in a similar way,
by comparing energies of two disjoint plaquettes to the
exact energies of two coupled plaquettes.
Small expectation values $\langle h_{ijk}\rangle$ relative to $\langle h_{ij}\rangle$
suggest rapid convergence of the cluster expansion.

Table \ref{tab:hijk} gives a summary of the ratios
$\langle h_{ij}\rangle/\langle h_{ijk}\rangle$
in the lowest states of the different sectors of the Hamiltonian.
The satisfactory convergence of the cluster expansion,
implies the integrity of bosonic states on the
lattice, at least for ladder geometry. Interestingly, it is still very good
for the Hubbard model with $U=6t$ where pair binding energy is positive on a plaquette.
This strengthens the argument that binding is generated dynamically on the lattice.
The holes remain tightly bound because correlated motion reduces their kinetic energy.

\begin{table}[h]
\begin{tabular}{c|cc}
     & S=0 & S=1  \\ \hline
  0 holes & 330 &  7.7  \\
  2 holes & 27 &  19.5  \\
\end{tabular}
\vskip0.4pc \caption{{\bf Convergence of the cluster expansion.}
The ratio $\langle h_{ij}\rangle/\langle h_{ijk}\rangle$ given for different
sectors in the Hamiltonian with $U=6t$ indicates excellent convergence of
CORE on a ladder.}
\label{tab:hijk}
\end{table}

\section{CORE calculation for the tight binding model}
\label{app:tightbinding}

In section \ref{sec:lts} we discussed the effects of breaking lattice translational symmetry
within the reduced Hilbert space. We argued that
interactions of increasing range gradually reduce the effects of symmetry breaking on the
spectrum. It is instructive to study this process in a simple model where a CORE
calculation can be carried easily to long ranges. Such an opportunity
is provided by the tight binding model on a chain:
\be
\cH=-\sum_{i}(c\yd_i c\nd_{i+1}+c\yd_{i+1} c\nd_i).
\label{tightbinding}
\ee
We apply CORE to the single electron sector of this model,
coarse graining it to blocks of 2 sites.
In each block we retain only the empty state $\ket 0_i$
and the single electron symmetric state of energy $-t$:
\be
f\yd_i\ket 0\equiv {1\over \sqrt{2}}(c\yd_{2i}+c\yd_{2i+1}).
\ee
Hence we can only hope to reconstruct the lowest of the 2 bands in the
folded Brillouin zone $k=[-\pi/2,\pi/2]$.

\begin{figure}[htb]
\begin{center}
\vspace*{13pt}
 \leavevmode \epsfxsize0.95\columnwidth
\epsfbox{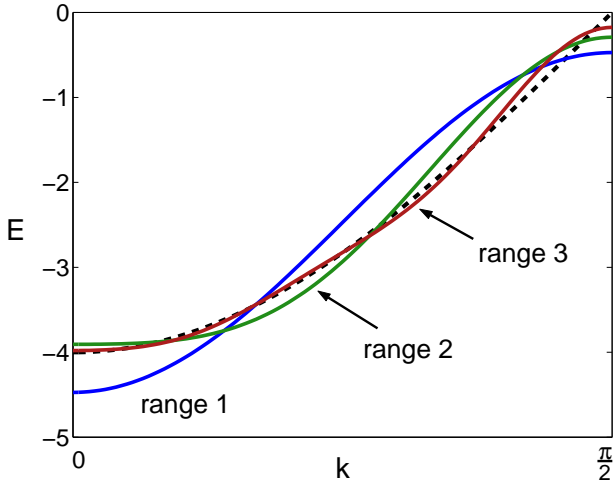}\vskip0.6pc
\caption{{\bf Tight binding dispersion from a CORE calculation.} The model is coarse
grained into dimers, and the cluster expansion is truncated at increasing hopping ranges.
Away from the dimer Brillouin zone edge, the approximate dispersion (solid lines)
converges rapidly to the exact solution (dashed line).
Its convergence is much slower at the dimer zone edge.}
\label{fig:tightbinding}
\end{center}
\end{figure}

The effective Hamiltonian generated
by CORE at any range of the cluster expansion is of the general form:
\be
\cH^{eff}=\sum_{ij}t_{ij}(f\yd_i f\nd_j+f\yd_j f\nd_i).
\ee
Such a Hamiltonian cannot reproduce the sharp band edge at $k=\pm\pi/2$ at any finite range
of hopping. However, as demonstrated in figure (\ref{fig:tightbinding}),
CORE calculations of increasing range introduce higher harmonics that
successively approximate the sharp edge. If one is interested in the properties of the
model far from the dimerized zone edge then by Fig.\ref{fig:tightbinding}
the effective Hamiltonian generated
by range-3 CORE should suffice.

\end{document}